\documentclass{jac}

\usepackage{amsfonts,amsmath,amssymb}
\usepackage[all]{xy}
\usepackage{stmaryrd}

\newcommand{\eqdf}{=_{\mathrm{df}}}

\newcommand{\ca}{\textsc{ca}}
\newcommand{\C}{\ensuremath{\mathcal{C}}}
\newcommand{\D}{\mathcal{D}}

\newcommand{\Id}{\mathrm{Id}}
\newcommand{\id}{\ensuremath{\mathrm{id}}}
\newcommand{\comp}{\circ}
\newcommand{\idD}{\mathrm{jd}}
\newcommand{\compD}{\bullet}

\newcommand{\eps}{\varepsilon}
\newcommand{\de}{\delta}
\newcommand{\dg}{\dagger}
\newcommand{\ddg}{\ddagger}

\newcommand{\restrict}[2]{#1|_{#2}}

\newcommand{\Set}{\ensuremath{\mathbf{Set}}}   
\newcommand{\Top}{\ensuremath{\mathbf{Top}}}   
\newcommand{\Unif}{\ensuremath{\mathbf{Unif}}} 

\newcommand{\coKl}{\ensuremath{\mathrm{coKl}}} 
\newcommand{\coEM}{\ensuremath{\mathrm{coEM}}}  

\begin{document}

\title{A Categorical Outlook on Cellular Automata}

\author{S. Capobianco}{Silvio Capobianco}

\author{T. Uustalu}{Tarmo Uustalu}

\address{Institute of Cybernetics at Tallinn University of Technology \\
Akadeemia tee 21\\ 12618 Tallinn, Estonia}
\email{{silvio, tarmo}@cs.ioc.ee}

\thanks{This research was supported by the European Regional
  Development Fund (ERDF) through the Estonian Center of Excellence in
  Computer Science (EXCS) and by the Estonian Science Foundation under
  grant no.~6940.}

\begin{abstract}\noindent
  In programming language semantics, it has proved to be fruitful to
  analyze context-dependent notions of computation, e.g., dataflow
  computation and attribute grammars, using comonads. We explore the
  viability and value of similar modeling of cellular automata.  We
  identify local behaviors of cellular automata with coKleisli maps of
  the exponent comonad on the category of uniform spaces and uniformly
  continuous functions and exploit this equivalence to conclude some
  standard results about cellular automata as instances of basic
  category-theoretic generalities. In particular, we recover
  Ceccherini-Silberstein and Coornaert's version of the Curtis-Hedlund
  theorem.
\end{abstract}

\maketitle

\section{Introduction}

Since the seminal work of Moggi~\cite{m91}, it has become standard in
programming language semantics to analyze functions producing effects
such as exceptions, input, output, interactive input-output,
nondeterminism, probabilistic choice, state, continuations using
monads. Specifically, effectful functions are identified with Kleisli
maps of a suitable monad on the category of pure functions.
Wadler~\cite{w92} put this view to further use in programming
methodology by extracting from it a very useful programming idiom for
purely functional languages like Haskell.

The dual view of context-dependent functions as coKleisli maps of a
comonad is equally useful, but less well known. Brookes and Geva
\cite{bg92} explained the ``intensional'' aspect of denotational
semantics in terms of the $\omega$-chain comonad on the category of
$\omega$-cpos. More recently, Uustalu and Vene \cite{uv06,uv07,uv08}
employed comonads to analyze dataflow computation and attribute
grammars and Hasuo et al.~\cite{hju07} treated tree transducers.

Characteristic of many context-dependent notions of computation is
shape-preserving transformation of some datastructure based on a value
update rule which is local in nature and applied uniformly to every
node.  This is the case with dataflow computation where such a
transformation is applied to a list or to a stream with a
distinguished position and with attribute grammars where computation
happens on a tree or a tree with a distinguished position (a
``zipper''). Cellular automata are similar, the datastructure being
the integer line or plane or, more generally, a group. It should
therefore be worthwhile to test the slogan that context-dependent
computation is comonadic also on cellular automata. To a degree, this
has already been done, as Piponi~\cite{p06} programmed cellular
automata in Haskell using the comonadic interface. However, he did not
use his modeling of cellular automata to prove properties about them
and also dropped the classical requirement that cellular automata rely
on a finite neighborhood only.

In this paper, we study the comonadic aspect of cellular automata
deeper. We identify cellular automata (more exactly their local
behaviors) with coKleisli maps of the exponent monad on $\Unif$, the
category of uniform spaces and uniformly continuous functions, and
explore whether this view can be useful. We see that it is: we can
conclude some standard results about cellular automata as instances of
category-theoretic generalities. In particular, we recover the
Curtis-Hedlund theorem~\cite{h69}---a characterization of global
behaviors of cellular automata---in the version of
Ceccherini-Silberstein and Coornaert~\cite{csc08} (this applies to
general discrete alphabets rather than finite alphabets only). This
theorem turns out to be an instance of the basic category-theoretic
fact that the coKleisli category of a comonad is isomorphic to the
full subcategory of its co-Eilenberg-Moore category given by the
cofree coalgebras. We also show that the comonadic view allows one to
see 2-dimensional cellular automata as 1-dimensional and treat
point-dependent cellular automata.

The paper is organized as follows. Section~\ref{sec:com} is a quick
introduction to comonads while Section~\ref{sec:top} reviews some
preliminaries about topological and uniform spaces.  In
Section~\ref{sec:ca}, we show that cellular automata local behaviors
are the same as coKleisli maps of a certain comonad.  In
Section~\ref{sec:ch}, we recover the Curtis-Hedlund theorem (in the
version of Ceccherini-Silberstein and Coornaert).  In
Section~\ref{sec:rev}, we reprove the reversibility principle. In
Sections~\ref{sec:2d}, \ref{sec:dep}, we discuss some further
applications of the comonadic view: 2-dimensional cellular automata as
1-dimensional and point-dependent cellular automata.

The paper assumes knowledge of basic category theory (categories,
functors, natural transformations, Cartesian closed categories), but
is self-contained in regards to comonads. For background material on
category theory and (co)monads in particular, we refer the reader to
Barr and Wells~\cite[Ch.~1, 3]{bw83}. We also assume the basics of
\ca\ as presented by Ceccherini-Silberstein and
Coornaert~\cite[Ch.~1]{csc10}.


\section{Comonads}
\label{sec:com}

Given two categories $\C$, $\D$ and a functor $L : \D \to \C$, a
\emph{right adjoint} to $L$ is given by a functor $R : \C \to \D$ and
two natural transformations $\eps : LR \to \Id_\C$ (the \emph{counit})
and $\eta : \Id_\D \to RL$ (the \emph{unit}) such that the diagrams
\[
\xymatrix@R=1.8pc{
L \ar[r]^{L \eta} \ar@{=}[dr]
  & LRL \ar[d]^{\eps L} \\
  & L 
}
\hspace*{1cm}
\xymatrix@R=1.8pc{
R \ar[d]_{\eta R} \ar@{=}[dr] 
  &  \\
RLR \ar[r]_{R \eps} 
  & R
}
\]
commute. Equivalently, a right adjoint may be given by an object
mapping $R : |\C| \to |\D|$, for any object $A \in |\C|$, a map
$\eps_A : LRA \to A$, and, for any objects $A \in |\D|$, $B \in |\C|$
and map $k : LA \to B$, a map $k^\ddg : A \to RB$ (the \emph{right
  transpose}) such that
\begin{itemize}
\item for any objects $A \in |\D|$, $B \in |\C|$, and map $k : LA \to
  B$, $\eps_B \comp L k^\ddg = k$, 
\item for any object $A \in |\C|$, $(\eps_A)^\ddg = \id_{R A}$,
\item for any objects $A, B \in |\D|$, $C \in |\C|$ and maps $f : A
  \to B$, $k : L B \to C$, $(k \comp L f)^\ddg = k^\ddg \comp f$.
\end{itemize}
The morphism mapping part of $R$ and unit $\eta$ define the right
transpose of $k : LA \to B$ by $k^\ddg \eqdf R k \comp \eta_A$. The
right transpose $(-)^\ddg$ determines the morphism mapping part of $R$
and unit $\eta$ by $R f \eqdf (f \comp \eps_A)^\ddg$, for $f : A \to
B$, and $\eta_A \eqdf (\id_{LA})^\ddg$.

A \emph{comonad} on a category $\C$ is given by a functor $D : \C \to
\C$ and natural transformations $\eps : D \to \Id_\C$ (the \emph{counit})
and $\de : D \to DD$ (the \emph{comultiplication}) making the diagrams
\[
\xymatrix@R=1.8pc{
D \ar[r]^{\de} \ar[d]_{\de} \ar@{=}[rd]
  & DD \ar[d]^{\eps D} \\
DD \ar[r]^{D\eps} & D
}
\hspace*{1cm}
\xymatrix@R=1.8pc{
D \ar[r]^-{\de} \ar[d]_{\de} & DD \ar[d]^{\de D} \\
DD \ar[r]^-{D\de} & DDD
}
\]
commute. Equivalently, a comonad can be given by an object mapping $D
: |\C| \to |\C|$, for any object $A \in |\C|$, a map $\eps_A : DA \to
A$, and, for any objects $A, B \in |\C|$ and map $k : DA \to B$, a map
$k^\dg : DA \to DB$ (the \emph{coKleisli extension}) such that
\begin{itemize}
\item for any objects $A, B \in |\C|$ and map $k : DA \to B$, $\eps_B
  \comp k^\dg = k$,
\item for any object $A$, $(\eps_A)^\dg = \id_{D A}$,
\item for any objects $A, B, C \in |\C|$ and maps $k : DA \to B$,
  $\ell : D B \to C$, $(\ell \comp k^\dg)^\dg = \ell^\dg \comp k^\dg$.
\end{itemize}
The morphism mapping part of $D$ and comultiplication $\de$ define the
coKleisli extension $(-)^\dg$ by $k^\dg \eqdf D k \comp \de_A$.
Conversely, the $(-)^\dg$ determines the morphism mapping part of $D$ and
comultiplication $\de$ by $D f \eqdf (f \comp \eps_A)^\dg$, $\de_A
\eqdf (\id_{D A})^\dg$.

A functor $L : \C \to \D$ with a right adjoint $(R, \eps, \eta)$,
defines a comonad on $\C$ with counit $\eps$ by $D \eqdf LR$, $\de
\eqdf L\eta R$, alternatively by $D A \eqdf L (R A)$, $k^\dg \eqdf L
k^\ddg$.

In the converse direction, a comonad $(D, \eps, \de)$ on $\C$ induces
a whole category of adjunctions $(\D, L, R, \eta)$ that have $\eps$ as
the counit and satisfy $D = LR$, $\de = L \eta R$, called
\emph{splittings} of the comonad.  This category has initial and final
objects, which are known as the coKleisli and coEilenberg-Moore
splittings of the comonad.

The \emph{coKleisli category} $\coKl(D)$ has as objects those of $\C$
and as maps from $A$ to $B$ those from $DA$ to $B$ of $\C$. The
identity $\idD_A$ on object $A$ is defined by $\idD_A \eqdf \eps_A$.
The composition $\ell \compD k$ of maps $k : DA \to B$ and $\ell : DB
\to C$ is $\ell \compD k \eqdf \ell \comp D k \comp \de_A = \ell
\comp k^\dg$. The functor $L : \coKl(D) \to \C$ in the coKleisli
splitting is defined by $L A \eqdf D A$, $L k \eqdf k^\dg$. The right
adjoint, unit and right transpose are defined by $R A \eqdf A$, $R f
\eqdf f \comp \eps_A$, $\eta_A \eqdf \id_{D A}$, $k^\ddg \eqdf k$.

The \emph{coEilenberg-Moore category} $\coEM(D)$ has as objects
coalgebras of $D$ and as maps coalgebra maps of $D$. A
\emph{coalgebra} of $D$ is given by an object $A \in |\C|$ and map $u
: A \to DA$ (the \emph{coalgebra structure}) making the diagrams
\[
\xymatrix@R=1.8pc{
A \ar[r]^{u} \ar@{=}[rd]
  & DA \ar[d]^{\eps_A} \\
  & A
}
\hspace*{1cm}
\xymatrix@R=1.8pc{
A \ar[r]^-{u} \ar[d]_{u} & DA \ar[d]^{\de_A} \\
DA \ar[r]^-{D u} & D(DA)
}
\]
commute. A \emph{coalgebra map}
between $(A, u)$ and $(B, v)$ is a map $f : A \to B$ making the
diagram
\[
\xymatrix@R=1.8pc{
A \ar[r]^-{u} \ar[d]_{f} 
  & DA \ar[d]^{D f} \\
B \ar[r]^-{v}
  & DB
}
\]
commute. The identity and composition are inherited from $\C$.  The
functor $L : \coEM(D) \to \C$ in the splitting of $D$ through
$\coEM(D)$ is the coalgebra-structure forgetful functor: $L (A, u)
\eqdf A$, $L f \eqdf f$. The right adjoint $R$ is defined by $R A
\eqdf (D A, \de_A)$, $R f \eqdf D f$. The unit and right transpose are
defined by $\eta_{(A,u)} \eqdf u$ and $k^\ddg \eqdf D k \comp u$ (for
$k : L (A, u) \to B)$.

The functor $R$ being the right adjoint of the forgetful functor
implies that, for any $B$, the coalgebra $R B = (D B, \de_B)$ is the
cofree coalgebra on $B$, i.e., for any coalgebra $(A, u)$, object $B$
and map $k : A \to B$, there is a unique map $f : A \to D B$, namely
$k^\ddg$, such that the diagrams 
\[
\xymatrix@R=1.8pc{
  & A \ar[dl]_{k} \ar[r]^-{u} \ar@{.>}[d]^{f} & DA \ar@{.>}[d]^{Df} \\
B & DB \ar[l]_{\eps_B} \ar[r]^-{\de_B} & D(DB)
}
\]
commute.

The unique splitting map between the coKleisli and coEilenberg-Moore
splitting is the functor $E : \coKl(D) \to \coEM(D)$ defined by $E A
\eqdf (D A, \de_A)$, $E k \eqdf D k \comp \de_A = k^\dg$. This functor
is a full embedding. The image of $E$ is the full subcategory of
$\coEM(D)$ given by the cofree coalgebras that is therefore isomorphic
to $\coKl(D)$.

A simple and instructive example of a comonad and its coKleisli and
coEilenberg-Moore splittings is given by the \emph{reader} (or
\emph{product}) \emph{comonad}. It is defined on any category $\C$
with finite products, but let us choose $\C$ to be $\Set$ (or $\Top$
or $\Unif$), so we can write pointwise definitions for intuitiveness.
Given some fixed object $C \in |\C|$, it is defined by $D A \eqdf A
\times C$, $D f (x, c) \eqdf (f (x), c)$, $\eps_A (x, c) \eqdf x$,
$\de_A (x, c) \eqdf ((x, c), c)$, $k^\dg (x, c) \eqdf (k (x, c), c)$.

The coKleisli category has as objects those of $\C$ and as maps from
$A$ to $B$ those from $A \times C$ to $B$. The identities and
composition are defined by $\idD (x, c) \eqdf x$, $(\ell \compD k)
(x, c) \eqdf \ell(k (x, c), c)$.

A coalgebra of $D$ is given by an object $A$ and a map $u : A \to A
\times C$ satisfying the laws of a coalgebra. Let us define
$(u_0(x), u_1(x)) \eqdf u(x)$. The laws impose that
$u_0(x) = x$ and $((u_0(x),u_1(x)), u_1(x)) =
(u(u_0(x)),u_1(x))$.  The first law defines $u_0$ and the
second becomes a tautology as soon as this definition is substituted
into it. Hence, a coalgebra is effectively the same as an object $A$
with an unconstrained map $u_1 : A \to C$.

A map between $D$-coalgebras $(A, u)$, $(B, v)$ is a map $f : A
\to B$ such that $(f(u_0(x)),u_1(x)) = (v_0(f(x)),
v_1(f(x)))$, which boils down to $u_1(x) = v_1(f(x))$.

The coEilenberg-Moore category has thus as objects pairs of an object
$A$ and map $u_1 : A \to C$ and a map between $(A, u_1)$, $(B, v_1)$
is map $f : A \to B$ such that $u_1(x) = v_1(f(x))$. The cofree
coalgebra on $A$ is the pair $(A \times C, {\delta_1}_A)$ where
${\delta_1}_A (x, c) \eqdf c$.

The isomorphism between $\coKl(D)$ and the category of cofree
$D$-coalgebras establishes a 1-1 correspondence between maps $k :
A \times C \to B$ and maps $f : A \times C \to B \times C$ such that
${\delta_1}_B (f (x, c)) = c$.


\section{Exponentials, topologies, and uniformities} 
\label{sec:top}

Given an object $C$ in a category $\C$ with finite products, it is
said to be \emph{exponentiable} if the functor $(-) \times C$ has a
right adjoint. This amounts to the existence, for any object $A$, of an
object $A^C$ (the \emph{exponential}) and map $\mathsf{ev}_A : A^C
\times A \to C$ (the \emph{evaluation}) as well as, for any
objects $A$, $B$ and map $k : A \times C \to B$, a map
$\mathsf{cur}(k) : A \to B^C$ (the \emph{currying} of $k$) satisfying
appropriate conditions. If every object of $\C$ is exponentiable, it is
called \emph{Cartesian closed}. Intuitively, exponentials are
internalized homsets. In $\Set$, every object $C$ is exponentiable and
the exponential $A^C$ is the set of all functions from $C$ to $A$.

Things are somewhat more complicated in the category $\Top$ of
topological spaces, as the exponential $A^C$ is to be the set of
continuous functions from $C$ to $A$, but it must also be given a
topology. Moreover, the evaluation $\mathsf{ev}_A : A^C \times A \to
C$ must be continuous and the currying $\mathsf{cur}(k) : A \to B^C$
of a continuous function must be continuous.

Not every topological space is exponentiable.
Hausdorff spaces are exponentiable
if and only if they are locally compact:
in this case, the exponential topology on the space of continuous functions from $C$ to $A$ is the \emph{compact-open topology}
generated by the sets $\{f : C \to A \mid f(K)\subseteq U\}$
with $K$ compact in $C$ and $U$ open in $A$~\cite{eh01,g85}.
In particular, discrete spaces are exponentiable
(which also follows from the discrete topology
making every function from it continuous)
and their compact-open topology is in fact the product topology.

That not all objects can act as exponents
is also true in the category $\Unif$ of uniform spaces
whose constituents we now define.
A \emph{uniform space} is a set $A$ endowed with a \emph{uniformity},
i.e., a collection $\mathcal{U}$
of binary relations on $A$
(called \emph{entourages})
satisfying the following properties:
\begin{enumerate}
\item
$\Delta\subseteq U$ for every $U\in\mathcal{U}$,
where $\Delta=\{(x,x)\mid x\in A\}$
is the \emph{diagonal}.
\item
If
$U\subseteq V$
and
$U\in\mathcal{U}$
then $V\in\mathcal{U}$.
\item
If $U,V\in\mathcal{U}$ then $U\cap V\in\mathcal{U}$.
\item
If $U\in\mathcal{U}$ then $U^{-1}\in\mathcal{U}$.
\item
If $U\in\mathcal{U}$ then
\begin{math}
V^2=\left\{
(x,y) \mid \exists z \mid (x,z),(z,y) \in V
\right\}\subseteq U
\end{math}
for some $V\in\mathcal{U}$.
\end{enumerate}
The simplest non-trivial uniformity on $A$
is the \emph{discrete uniformity},
made of all the supersets of the diagonal.
A uniformity
induces a topology
as follows:
$\Omega \subseteq A$ is open if and only if, for every $x \in \Omega$,
there exists $U \in \mathcal{U}$ such that
\begin{math}
\{y\in A\mid(x,y)\in U\} \subseteq \Omega.
\end{math}
Such topology is Hausdorff if and only if
\begin{math}
\bigcap_{U\in\mathcal{U}}U=\Delta.
\end{math}
The discrete uniformity induces the discrete topology,
but is not the only one that does (cf.~\cite[I-5]{i64}),
i.e., uniform spaces may be discrete
without being \emph{uniformly discrete}.

A map $f : A \to B$ between uniform spaces
is \emph{uniformly continuous} (briefly, u.c.)\
if, for every entourage $V$ on $B$,
there is an entourage $U$ on $A$ such that $(f \times f)(U) \subseteq V$.
Any u.c.\ function is continuous
in the topology induced by the uniformities:
the converse is true if $A$ is compact~\cite[II-24]{i64}
but false in general even for metric spaces.
The \emph{product uniformity} is the coarsest uniformity
that makes the projections uniformly continuous:
the topology induced by the product uniformity is the product topology.
A product of discrete uniformities is called \emph{prodiscrete}.

In $\Unif$, uniformly discrete objects are exponentiable \cite[III.19
and III.21]{i64}. Again, the reason is that every function from $C$ is
u.c.\ as soon as $C$ is uniformly discrete.


\section{Cellular automata as coKleisli maps}
\label{sec:ca}

Classically, a \emph{cellular automaton} on a monoid $(G, 1_G, \cdot)$
(the \emph{universe})\footnote{Instead of the monoid, one usually
  takes a group in the cellular automata literature. But we do not
  need inverses in this paper.} and set $A$ (the
\emph{alphabet})\footnote{The alphabet is
  often required to be finite. We make this assumption only where we need it.} is given by a finite subset $N$ of $G$ (the
\emph{support neighborhood}) and function $d : A^N\to A$
(the \emph{transition rule}).

Any cellular automaton induces a \emph{local behavior} $k : A^G \to
A$ via $k(c) \eqdf d (\restrict{c}{N})$.

One speaks of elements $A^N$ of finite subsets $N \subseteq
G$ as \emph{patterns} and elements of $A^G$ as \emph{configurations}.
Transition rules work on patterns, local behaviors on configurations.
Cellular automata that induce the same local behavior are considered
equivalent. In this paper, we will not distinguish between equivalent
cellular automata, hence we can identify cellular automata with their
local behaviors.

If the alphabet $A$ is finite, a function $k : A^G \to A$ is a local
behavior (i.e., the local behavior of some cellular automaton) if and
only if it is continuous for the discrete topology on $A$ and the
product topology on $A^G$.

In the general case (where $A$ may be infinite), the above equivalence
does not generally hold, but a refinement does. A function $k : A^G
\to A$ is then a local behavior iff it is uniformly continuous for the
discrete uniformity on $A$ and product uniformity on $A^G$.\footnote{
(Cf.~\cite[Th. 1.9.1]{csc10})
Every entourage of the prodiscrete uniformity
contains an entourage of the form
\begin{math}
V_N = \left\{
(c,e)
\mid \restrict{c}{N} = \restrict{e}{N}
\right\}
\end{math}
with $N \subseteq G$ finite.
If $k : A^G \to A$ is u.c.\ with $A$ uniformly discrete, then
$(k \times k)(V_N) \subseteq \Delta$ for some finite $N \subseteq G$:
thus, $k(c)$ only depends on $\restrict{c}{N}$.
}
The finite case becomes an instance: if $A$ is finite, then $A^G$ is compact and therefore any
continuous function $k : A^G \to A$ is uniformly continuous.

Based on these observations, we henceforth take it as a definition
that a local behavior on a set (the alphabet) $A$ is a uniformly
continuous function $k : A^G \to A$ wrt.\ the prodiscrete uniformity on
$A^G$ and forget about the definition of cellular automata in terms of
a support neighborhood and a transition rule.

Any local behavior $k$ induces a \emph{global behavior} $k^\dg : A^G
\to A^G$, a map between configurations, via
$
k^\dg (c)(x) \eqdf k (c \rhd x)
$
where $\rhd : A^G \times G \to A^G$ (the \emph{translation} of
configurations) is defined by
$
(c \rhd x)(y) \eqdf c(x \cdot y)
$
The translation is a uniformly continuous function. It follows that
the global behavior $k^\dg$ is also uniformly continuous.

Local behaviors on a fixed universe $G$ and fixed alphabet $A$ form a
monoid with unit $\idD$ given by $\idD (c) \eqdf c (1_G)$ and
multiplication $\bullet$ given by $\ell \compD k \eqdf \ell \comp
k^\dg$. Indeed, it is easy to see that $\idD$ is uniformly continuous
and $\compD$ preserves uniform continuity (because $(-)^\dg$ does) and
the monoid laws turn out to hold too.

We now make two small generalizations and make a richer category out
of local behaviors: after all, a monoid is a category with one
object. First, we do not insist that the alphabet be a discrete
uniform space, it may be any uniform space.  And second, we give up
the idea of a fixed alphabet: we let the local behavior change the
alphabet.

For a fixed monoid $G$ (the universe), we redefine a \emph{local
  behavior} between two general uniform spaces (the source and target
alphabets) $A$ and $B$ to be a uniformly continuous function $k : A^G
\to B$ where $A^G$ is given the product uniformity.

Local behaviors now make a category that has as objects alphabets and
as maps local behaviors between them.  The identity on $A$ is $\idD_A
: A^G \to A$ given by $\idD_A (c) \eqdf c (1_G)$ and the composition
$\ell \compD k : A^G \to C$ of two maps $k : A^G \to B$ and $\ell :
B^G \to C$ given by $\ell \compD k \eqdf \ell \comp k^\dg$ where
$k^\dg : A^G \to B^G$ is defined by $k^\dg(c)(x) \eqdf k(c \rhd_A x)$
from $\rhd_A : A^G \times G \to A^G$ defined by $(c \rhd_A x)(y) \eqdf
c(x \cdot y)$.  Notice that these definitions coincide exactly with
those we made for the monoid of local behaviors above, except that
local behaviors can now mediate between different alphabets that need
not be uniformly discrete. The function $\idD_A$ is still uniformly
continuous for any $A$ and the operation $\compD$ preserves uniform
continuity.

While the generalized definition of local behaviors is more liberal
than the classical one, it is conservative over it in the following
sense: The local behaviors from any uniformly discrete space $A$ back
to itself are exactly the classical local behaviors on $A$ seen as a
set.

We will now recover our category of local behaviors from a categorical
generality, by showing that it is a straightforward instance of the
coKleisli construction for a comonad.

Any fixed monoid $(G, 1_G, \cdot)$ determines a comonad $(D, \eps,
\delta)$ on $\Unif$ (in fact, on any category where the carrier $G$ is
exponentiable, so also, e.g., on $\Set$ and $\Top$) (the
\emph{cellular automata} or \emph{exponent} \emph{comonad}) as
follows. The object mapping part of $D$ is defined by $D A \eqdf A^G$,
where $A^G$ is the $G$-exponential of $A$, i.e., the space of uniformly
continuous functions from $G$ to $A$ equipped with the prodiscrete
uniformity.  The morphism mapping part is defined by $D f \eqdf f^G$,
i.e., $D f (c) \eqdf f \comp c$.  The components of the counit $\eps_A
: A^G \to A$ and comultiplication $\de_A : A^G \to (A^G)^G$ are
defined by $\eps_A (c) \eqdf c(1_G)$ and $\de_A (c)(x) \eqdf c \rhd_A
x$ (so that $\de_A (c) (x) (y) = c(x \cdot y)$); these functions are
uniformly continuous. The general definition of the coKleisli
extension $(-)^\dg$ via the morphism mapping part of $D$ and
comultiplication $\de$ tells us that $k^\dg (c)(x) = D k (\de_A(c))(x)
= k (\de_A(c)(x)) = k (c \rhd_A x)$.

The laws of a comonad are proved from the monoid laws for $G$ by the
following calculations (we omit the proofs of the naturality
conditions of $\eps$ and $\de$).
\[
\begin{array}{c}
\eps_{DA} (\de_A (c)) (x) 
= \de_A(c) (1_G) (x) 
= c (1_G \cdot x) 
= c(x) 
\\[1ex]
c (x)
= c (x \cdot 1_G)
= \de_A (c) (x) (1_G)
= \eps_A (\de_A (c) (x))
= D\eps_A (\de_A (c)) (x)
\\[1ex]
\de_{DA} (\de_A (c)) (x) (y) (z) 
= \de_A(c) (x \cdot y) (z) 
= c ((x \cdot y) \cdot z) 
\hspace*{5cm} \\ \hspace*{1cm} 
= c (x \cdot (y \cdot z)) 
= \de_A (c) (x) (y \cdot z) 
= \de_A (\de_A (c) (x)) (y) (z)
= D\de_A (\de_A (c)) (x) (y) (z)
\end{array}
\]

As we have seen, a comonad on a category always defines two canonical
splittings of its underlying functor into two adjoint functors. The
coKleisli splitting of our comonad $D$ on $\Unif$ goes via the
coKleisli category which has as objects those of $\Unif$ and as maps
from $A$ to $B$ those from $D A$ to $B$ in $\Unif$. The identity on
$A$ is $\idD_A \eqdf \eps_A$ and the composition of $k$ and $\ell$ is
$\ell \compD k \eqdf \ell \comp k^\dg$. Note that these are exactly
the data of the category of local behaviors that we introduced above.
But this time we do not have to prove that the unital and
associativity laws of the category hold. Our proof obligations went
into establishing that the comonad data are well defined and the comonad
laws hold.


\section{Retrieving the Curtis-Hedlund theorem} 
\label{sec:ch}

Let $(D, \eps, \de)$ be the $G$-exponential comonad on $\Unif$ for a
given monoid $(G, 1_G, \cdot)$, with $G$ endowed with the discrete
uniformity, as introduced in the previous section.

As we know from Section~\ref{sec:com},
$\coKl(D)$ is equivalent to the category of cofree $D$-coalgebras
under a comparison functor $E$ that sends a coKleisli map (local
behavior) $k : D A \to B$ to the cofree coalgebra map $k^\dg : (D A,
\de_A) \to (D B, \de_B)$, which, as a map of $\Unif$, we know to be
the corresponding global behavior.
Hence, a map $f : D A \to D B$ would be a global behavior if and only
if $f$ is a cofree coalgebra map.

Now, given an arbitrary comonad, it is usually of interest to study
its general coalgebras and not only the cofree ones.  We too
follow this thumb rule.

By definition, objects in $\coEM(D)$ are pairs of objects $A$ and maps
$u:A\to DA$ in $\Unif$ satisfying
\[
\xymatrix@R=1.2pc{
A \ar[r]^-{u} \ar@{=}[rd] & A^G \ar[d]^{\eps_A} \\
& A
}
\hspace*{1cm}
\xymatrix@R=1.2pc{
A \ar[r]^-{u} \ar[d]_{u} & A^G \ar[d]^{\de_A} \\
A^G \ar[r]^-{u^G} & (A^G)^G
}
\]
In our case,
the first equation simply means
\begin{math}
u(a)(1_G)=a 
\end{math}
while the second one simplifies to 
\begin{math}
u(a)(x\cdot y)=u(u(a)(x))(y).
\end{math}
This writing, however, is cumbersome and unexplicative.

To see more, we uncurry
\begin{math}
u:A\to A^G
\end{math}
to 
\begin{math}
\otimes:A\times G\to A,
\end{math}
so that 
\begin{math}
a \otimes x=u(a)(x).
\end{math}
Then the two equations become
\begin{math}
a\otimes 1_G=a,
\end{math}
and
\begin{math}
a\otimes (x\cdot y)=(a\otimes x)\otimes y.
\end{math}
Diagrammatically, this is to require commutation of
\[
\xymatrix@R=1.2pc{
A \ar[r]^-{\rho_A} \ar@{=}[rrd]
  & A \times 1 \ar[r]^-{A \times 1_G} & A \times G \ar[d]^{\otimes} \\
& & A
}
\hspace*{1cm}
\xymatrix@R=1.2pc{
(A\times G) \times G \ar[r]^-{\alpha_{A,G,G}} \ar[d]_{\otimes \times G}
 & A\times (G \times G) \ar[r]^-{A \times (\cdot)} & A \times G \ar[d]^{\otimes} \\
A \times G \ar[rr]^-{\otimes} & & A
}
\]
where $\rho$ and $\alpha$ are the right unital and associative laws of
the product monoidal structure.
But these are precisely the laws of a \emph{(right) action} of $G$ on
$A$ (where $A$ is a uniform space, so we expect an action to also be
uniformly continuous).

Let us now consider coalgebra maps.
A map
\begin{math}
f:(A,u)\to(B,v)
\end{math}
in $\coEM(D)$ is a map in $\Unif$ that commutes with $u$ and $v$ or
(which is equivalent) with their uncurried forms $\otimes$ and
$\oslash$ as shown on the left and right diagrams below, respectively:
\[
\xymatrix@R=1.2pc{
A \ar[r]^-{u} \ar[d]_{f} & A^G \ar[d]^{f^G} \\
B \ar[r]^-{v} & B^G
}
\hspace*{1cm}
\xymatrix@R=1.2pc{
A\times G \ar[r]^-{\otimes} \ar[d]_{f\times G} & A \ar[d]^{f} \\
B\times G \ar[r]^-{\oslash} & B
}
\]
Clearly, coalgebra maps are just action maps. 

We are now ready to consider maps of cofree $D$-coalgebras. The
uncurried form of $\de_A : A^G \to (A^G)^G$ is $\rhd_A : A^G \times G
\to A^G$. By what we have shown, a map $f : A^G \to B^G$ in $\Unif$ is
a map between $(A^G, \de_A)$ and $(B^G, \de_B)$ if and only if $f (c
\rhd_A x) = (f (c) \rhd_B x)$. We see that maps between the cofree
coalgebras $(A^G, \de_A)$ and $(B^G, \de_B)$ are precisely those maps
$f : A^G \to B^G$ that commute with the translation!

With our reasoning, we have reproved the version of Curtis-Hedlund
theorem given by Ceccherini-Silberstein and Coornaert
(\cite[Th.~1.1]{csc08}, \cite[Th.~1.9.1]{csc10}):
global behaviors between uniform spaces $A$ and $B$ are those uniformly
continuous functions between the product uniformities on $A^G$ and
$B^G$ that commute with the translation.
Keep in mind that we are allowing arbitrary uniform spaces as alphabets,
and speaking of global behaviors in the sense of Section~\ref{sec:ca}.
It is, however, immediate to specialize to the statement of C.-S. and C. 
by requiring $A$ and $B$ to be uniformly discrete.
Then local behaviors are precisely
the finitary functions from $A^G$ to $B$.

The original Curtis-Hedlund theorem (\cite{h69},
\cite[Th.~1.8.1]{csc10}) corresponds to the special case where $A$ is
finite and discrete. In this case, $A^G$ is compact and
any continuous function between $A^G$ and $B^G$ is then uniformly
continuous. We conclude, in this case, that global behaviors between
$A$ to $B$ are those continuous functions between the product
topologies on $A^G$ and $B^G$ that commute with the translation.

We have seen that the cofree coalgebra $(B^G, \delta_B : B^G \to
(B^G)^G)$ on $B$ is the currying of the translation action $(B^G,
\rhd_B : B^G \times G \to B^G)$. Because of the cofreeness on $B$ of
this coalgebra, the action must also be the cofree.

In basic terms, this means that for any action $(A, \otimes : A \times
G \to A)$ and map $p : A \to B$, there is a unique action morphism $f$
to the translation action such that $\eps_B \circ f = p$, i.e., a map
$f : A \to B^G$ such that the following diagrams commute:
\[
\xymatrix@R=1.2pc{
A \times G \ar[r]^-{\otimes} \ar@{.>}[d]_{f \times G}
 & A \ar[dr]^{p} \ar@{.>}[d]^{f}
   & \\
B^G \times G \ar[r]^-{\rhd_B}
 & B^G \ar[r]^{\eps_B}
   & B 
}
\]

This fact can of course be proved from first principles, but we
learned it for free!


\section{Retrieving the reversibility principle} 
\label{sec:rev}

We will now recover the reversibility principle. Let again $(G, 1_G,
\cdot)$ be a monoid and $(D, \eps, \de)$ be the $G$-exponential
comonad on $\Unif$ as introduced in Section~\ref{sec:ca}.

Suppose we have a cofree coalgebra map $f : (D A, \de_A) \to (D B,
\de_B)$ so that $f$ has an inverse $f^{-1}$ as a map of $\Unif$.

A very simple diagram chase (not specific to our particular comonad;
we only use that $D$ is a functor!) shows that $f^{-1}$ is also a
cofree coalgebra map, i.e., an inverse of $f$ in $\coEM(D)$:
\[
\xymatrix@R=1.2pc{
  & DA \ar[r]^-{\de_A} \ar[d]_{f}
    & D(DA) \ar[d]^{D f}  \ar@{=}[r]
      & D (DA)\\
D B \ar[ru]^{f^{-1}} \ar@{=}[r]  
  & DB \ar[r]^-{\de_B} 
    & D(DB) \ar[ru]_{Df^{-1}} 
      & 
}
\]

We have thus reproved the reversibility principle
\cite[Th.~1.10.1]{csc10}: A global behavior $f : A^G \to B^G$ (a
uniformly continuous function commuting with the translation) between
uniform spaces $A$, $B$ is reversible if $f$ has a uniformly
continuous inverse. 

If both $A$ and $B$ are finite and discrete, an inverse $f^{-1}$ of
$f$ is necessarily uniformly continuous, because, in this
  case, $A^G$ and $B^G$ are compact Hausdorff and $f^{-1}$ is
  continuous. So we obtain a special case \cite[Th.~1.10.2]{csc10}:
A global behavior $f : A^G \to B^G$ (a continuous function commuting
with the translation) between uniform spaces $A$, $B$ is reversible if
$f$ has an inverse.

In general, an inverse of a uniformly continuous function is not
necessarily uniformly continuous.  For a counterexample,
see~\cite[Example 1.10.3]{csc10}.

For reversibility, it is useful, if the monoid $G$ is actually a
group.  In particular, for reversibility of $\delta_A : D A \to D(D
A)$, $G$ must be a group.


\section{Distributive laws and 2-dimensional cellular automata}
\label{sec:2d}

We will now proceed to two variations on the theme of cellular
automata as co\-Kleisli maps---2-dimensional (classical) cellular
automata and point-dependent cellular automata. In both cases we first
introduce some further comonad theory relevant for our cause.

Sometimes, but not always, the composition $D^1D^0$ of two comonads
$D^0$ and $D^1$ on the same category $\C$ is a comonad. It is the
case, if there is a distributive law of $D^1$ over $D^0$. A
\emph{distributive law} of a comonad $(D^1, \eps^1, \de^1)$ over a
comonad $(D^0, \eps^0, \de^0)$ is a natural transformation $\kappa :
D^1D^0 \to D^0D^1$ making the diagrams 
\[
\xymatrix@R=1.2pc@C=1pc{
D^1D^0 \ar[rr]^{\kappa} \ar[dr]_{D^1 \eps^0} 
  & & D^0D^1  \ar[dl]^{\eps^0 D^1} \\
& D^1 & 
}
\hspace*{1cm}
\xymatrix@R=1.2pc{
D^1D^0 \ar[rr]^-{\kappa} \ar[d]_{D^1 \de^0} 
  & & D^0D^1  \ar[d]^{\de^0 D^1} \\
D^1D^0D^0 \ar[r]^-{\kappa D^0}
  & D^0D^1D^0 \ar[r]^-{D^0 \kappa}
    & D^0D^0D^1
}
\]
\[
\xymatrix@R=1.2pc@C=1pc{
D^1D^0 \ar[rr]^{\kappa} \ar[dr]_{\eps^1 D^0} 
  & & D^0D^1 \ar[dl]^{D^0 \eps^1} \\
& D^0 & 
}
\hspace*{1cm}
\xymatrix@R=1.2pc{
D^1D^0 \ar[rr]^-{\kappa} \ar[d]_{\de^1 D^0} 
  & & D^0D^1  \ar[d]^{D^0 \de^1} \\
D^1D^1D^0 \ar[r]^-{D^1 \kappa}
  & D^1D^0D^1 \ar[r]^-{\kappa D^1}
    & D^0D^1D^1
}
\]
commute. A distributive law induces a comonad $(D, \eps, \de)$ defined
by $D \eqdf D^1D^0$, $\eps \eqdf \eps^1\eps^0$, $\de \eqdf
D^1 \kappa D^0 \comp \de^1 \de^0$.

A distributive law also induces comonad liftings. For the lack of
space, we concentrate on the coKleisli side of the picture. Given a
distributive law $\kappa$ of $D^1$ over $D^0$, the comonad $D^0$ on
$\C$ lifts to $\coKl(D^1)$, i.e., induces a comonad $\bar{D^0}$ on
$\coKl(D^1)$.  This comonad is defined by: $\bar{D^0} A \eqdf D^0 A$,
$\bar{D^0} k \eqdf D^0 k \comp \kappa_A : D^1 D^0 A \to D^0 B$ (for $k :
D^1 A \to B$), $\eps_A \eqdf \eps^0_A \comp \eps^1_{D^0 A}$, $\de_A
\eqdf \de^0_A \comp \eps^1_{D^0 A}$.

Via this lifting, the coKleisli category of the composite comonad $D$
and the other ingredients of the coKleisli splitting of $D$ can be
obtained by a double coKleisli construction: We have $\coKl(D) =
\coKl(\bar{D^0})$ (equal strictly, not just isomorphic).

A simple example of a distributive law is obtained by taking $D^1$ to
be any comonad and $D^0$ the product comonad defined by $D^0 A \eqdf A
\times C$ for a fixed object $C \in |\C|$. The distributive law
$\kappa$ of $D^1$ over $D^0$ is given by $\kappa_A \eqdf \langle D^1
\pi_0, \eps^1 \comp D^1 \pi_1 \rangle : D^1 (A\times C) \to D^1 A
\times C$.  It follows that the functor $D$ defined by $D A \eqdf D^1
(A \times C)$ is a comonad.

Given now two monoids $G_0$, $G_1$, we can think of a map $k :
(A^{G_0})^{G_1} \to B$ in $\Unif$ as a ``2-dimensional'' (2D) cellular
automaton on the universes $G_0$, $G_1$ between alphabets $A$ and $B$
(relying on the isomorphism $A^{G_0 \times G_1} \cong
(A^{G_0})^{G_1}$).

Such a cellular automaton is by definition the same thing as a
``1-dimensional'' (1D) cellular automaton on the universe $G_1$
between alphabets $A^{G_0}$ and $B$. Note that we can only see 2D
cellular automata as 1D in this way, if we allow source and target
alphabets of a cellular automaton to differ and if we do not require
them to be uniformly discrete (notice that $A^{G_0}$ carries the
prodiscrete uniformity). But this view of 2D cellular automata as 1D,
although nice, suffers from a serious drawback.  Since the 1D views do
not have the same source alphabets as the 2D originals, they do not
compose the same way.

Distributive laws come to help. Let $D^0 A \eqdf A^{G_0}$ and $D^1 A
\eqdf A^{G_1}$. There is a distributive law $\kappa : D^1D^0 \to
D^0D^1$ defined by $\kappa_A (c) (x_1)(x_0) = c(x_0)(x_1) :
(A^{G_0})^{G_1} \to (A^{G_1})^{G_0}$. Hence, the functor $D A \eqdf
(A^{G_0})^{G_1}$ is a comonad, which is hardly a surprise. But there
is more: We know that $\coKl(D) = \coKl(\bar{D^0})$. Hence, a good
view of $k : (A^{G_0})^{G_1} \to B$ as a 1D cellular automaton is not
as a $\Unif$-cellular automaton on the universe $G_1$ between the
alphabets $A^{G_0}$ and $B$, but on as a $\coKl(D^1)$-cellular
automaton on the universe $G_0$ between the alphabets $A$ and $B$.
Then 1D views compose exactly as their 2D originals. We see that it
makes sense to consider maps of categories other than $\Unif$! 

1D views of 2D cellular automata were of interest to Dennuzio et
al.~\cite{df08}


\section{Comonad maps and point-dependent cellular automata}
\label{sec:dep}

To make a category out of comonads over a fixed category $\C$ one
needs a suitable notion of comonad maps. A \emph{comonad map}
between two comonads $(D, \eps, \de)$ and $(D', \eps', \de')$ on $\C$
is a natural transformation $\tau : D \to D'$ making the diagrams
\[
\xymatrix@R=0.35pc{
D \ar[rd]^-{\eps} \ar[dd]_{\tau} \\
  & \Id_\C \\
D' \ar[ru]_-{\eps'}
}
\hspace*{1cm}
\xymatrix@R=0.7pc{
D \ar[r]^{\de} \ar[dd]_{\tau}
  & DD \ar[dd]^{\tau \tau} \\
\\
D' \ar[r]^-{\de'} 
  & D'D'
}
\]
commute. Comonads and comonad maps on $\C$ form a category. The
identity and composition of comonad maps is inherited from the
category of natural transformations between endofunctors on $\C$.

A comonad map $\tau$ between $(D, \eps, \de)$ and $(D', \eps', \de')$
relates the coKleisli and coEilenberg-Moore categories between the two
comonads.  It defines a functor from $\coKl(D')$ to $\coKl(D)$ and a
functor from $\coEM(D)$ to $\coEM(D')$.

We now introduce point-dependent cellular automata (studied under the
name of non-uniform cellular automata by Cattaneo et
al.~\cite{cdfp09}). For a set $G$, the local behavior of a
point-dependent cellular automaton between uniform spaces $A$, $B$ is
a uniformly continuous function $k : A^G \times G \to B$. Note the
added second argument compared to the definition of a classical local
behavior.

It turns out that local and global behaviors of point-dependent
cellular automata can be analyzed in the same way as those of
classical cellular automata. In particular, their local behaviors are
the same thing as coKleisli maps of a suitable comonad $(D, \eps,
\de)$ on $\Unif$ and global behaviors are the corresponding cofree
coalgebra maps!

Let us review the data of the comonad. The object mapping of $D$ is
defined by $D A \eqdf A^G \times G$ and the morphism mapping by $D f
\eqdf f^G \times G$, i.e., $D f (c, x) \eqdf (\lambda y. f(c(y)), x)$.
The components of the counit and comultiplication $\eps_A : A^G \times
G \to A$ and $\de_A : A^G \times G \to (A^G \times G)^G \times G$ are
defined by $\eps_A (c, x) \eqdf c(x)$, $\de_A (c, x) = (\lambda y. (c,
y), x)$. Accordingly, the coKleisli extension $k^\dg : A^G \times G
\to B^G \times G$ of a map $k : A^G \times G \to B$ is forced to
satisfy $k^\dg (c, x) = D k (\de (c, x)) = D k (\lambda y. (c, y), x)
= (\lambda y. k (c, y), x)$.

When is a map $f : A^G \times G \to B^G \times G$ a global behavior?
It is a global behavior iff it is a cofree coalgebra map. Not
surprisingly at all, the conditions for $f$ being a cofree coalgebra
map reduce to the condition that $f (c, x) = (g(c), x)$ for some $g :
A^G \to B^G$.

Assume $G$ is endowed with a monoid structure $(1_G, \cdot)$. Let
$(D', \eps', \de')$ be the comonad of classical cellular automata.
The translation $\rhd$ is a comonad map from $D$ to $D'$. Accordingly,
any classical local behavior is also a point-dependent local behavior
that simply makes no use the point information that is available.


\section{Conclusions}
\label{sec:concl}

It was not the purpose of this paper to prove deep or difficult
theorems.  Rather, we set out to experiment with definitions. We deem
that this experiment succeeded. We were pleased to learn that, from
the category-theoretic point-of-view, cellular automata are a
``natural'' construction with ``natural'' properties. Crucially,
classical cellular automata are coKleisli maps of the exponential
comonad on $\Unif$, and it is harmless to accept alphabets with
nondiscrete uniformities and variation of alphabets, once it has been
decided that local behaviors are uniformly continuous functions. But
other base categories can be useful too, as the example of
2-dimensional cellular automata as 1-dimensional shows.

We hope to be able to extend this work to cover more results of
cellular automata theory, in particular results toward the Garden of
Eden theorem.

\subsubsection*{Acknowledgments}

We are grateful to Jarkko Kari and Pierre Guillon for comments.


\end{document}